\documentclass[conference]{IEEEtran}
\IEEEoverridecommandlockouts

\usepackage{cite}
\usepackage{amsmath,amssymb,amsfonts}
\usepackage{algorithmic}
\usepackage{graphicx}
\usepackage{textcomp}
\usepackage{xcolor}
\usepackage{tikz}
\usepackage{color,soul}
\usepackage{float}
\usetikzlibrary{arrows.meta,positioning,fit,calc,shapes.multipart,backgrounds}

\def\BibTeX{{\rm B\kern-.05em{\sc i\kern-.025em b}\kern-.08em
    T\kern-.1667em\lower.7ex\hbox{E}\kern-.125emX}}
\begin{document}

\title{BR-FiLM: Bounded Residual Channel-Quality Conditioning for Automatic Modulation Recognition}

\author{\IEEEauthorblockN{Tahmid Zaman Tahi}
\IEEEauthorblockA{\textit{Electrical and Computer Engineering} \\
\textit{Virginia Commonwealth University}\\
Richmond, VA, USA\\
tahit@vcu.edu}
\and
\IEEEauthorblockN{Syed Samiul Alam}
\IEEEauthorblockA{\textit{Electrical and Computer Engineering} \\
\textit{Virginia Commonwealth University}\\
Richmond, VA, USA\\
alams10@vcu.edu}
\and
\IEEEauthorblockN{Haolin Tang}
\IEEEauthorblockA{\textit{School of Engineering and Computing} \\
\textit{Fairfield University}\\
Fairfield, CT, USA \\
htang@fairfield.edu}
\and
\IEEEauthorblockN{Yanxiao Zhao}
\IEEEauthorblockA{\textit{Electrical and Computer Engineering} \\
\textit{Virginia Commonwealth University}\\
Richmond, VA, USA\\
yzhao7@vcu.edu}
\and
\IEEEauthorblockN{Jun Huang}
\IEEEauthorblockA{\textit{Electrical Engineering/Computer Science} \\
\textit{South Dakota State University}\\
Brookings, SD, USA \\
jun.huang@sdstate.edu}
\and
\IEEEauthorblockN{Min Song}
\IEEEauthorblockA{\textit{Electrical and Computer Engineering} \\
\textit{Stevens Institute of Technology}\\
Hoboken, NJ, USA \\
msong6@stevens.edu}
}

\maketitle
\begin{abstract}
Automatic Modulation Recognition (AMR) plays a crucial role in enabling robust, adaptive, and secure communication for military and civilian applications. Deep learning has enabled effective AMR methods that overcome the computational inefficiency of traditional approaches. However, these deep learning based methods often degrade significantly in low SNR conditions, where noise obscures modulation-discriminative waveform features. In this paper, we propose Bounded Residual Feature-wise Linear Modulation (BR-FiLM), a channel-quality conditioning block for AMR, which can be inserted into AMR classifiers with intermediate feature representations. We construct BR-FiLMNet by inserting the proposed BR-FiLM block into a Multi-Channel Convolutional Long Short-Term Deep Neural Network (MCLDNN) backbone. BR-FiLMNet conditions convolutional, recurrent, and dense features through gated residual corrections while preserving the original I/Q-driven feature path. Experimental results on RadioML 2016.10a show that BR-FiLMNet improves mean accuracy from 61.79\% to 67.74\% and low-SNR accuracy (SNR $\leq$ 0 dB) from 37.12\% to 46.44\% compared to MCLDNN. We further evaluate BR-FiLMNet against recent transformer-style baselines. The results indicate that BR-FiLMNet delivers significant and reliable performance gains, as validated through paired statistical testing.
\end{abstract}

\section{Introduction}

Automatic Modulation Recognition (AMR) is the process of automatically identifying the modulation scheme of a received signal. It plays a crucial role in modern military and civilian communications by enabling receivers to recognize unknown signals without prior transmitter information. This capability improves situational awareness, supports signal intelligence, and enhances communication resilience in complex and contested spectrum environments.

Traditional AMR methods rely mainly on likelihood-based inference or handcrafted signal features, but these approaches often require prior parameter knowledge, high computational cost, or expert-designed descriptors~\cite{zhang2022deep}. Deep Learning (DL)-based AMR has therefore become dominant because neural networks can learn discriminative representations directly from raw In-phase/Quadrature (I/Q) samples~\cite{zhang2022deep}. Convolutional Neural Network (CNN)-based, Recurrent Neural Network (RNN)-based, and hybrid CNN-RNN models have been widely studied for capturing local and temporal signal structures~\cite{hong2017automatic,xu2020spatiotemporal}.

Despite these advances, DL-based AMR models remain highly vulnerable in low Signal-to-Noise Ratio (SNR) environments, especially below 0 dB. 
In this regime, modulation-specific phase, amplitude, and temporal patterns are partially obscured, making the learned representation less reliable. A common response is to suppress noise before classification using a denoising front-end. However, waveform-level denoising can also remove fine modulation-discriminative structure, particularly for signals whose identity depends on subtle phase, amplitude, or envelope variations. This motivates a different design principle: rather than reconstructing the received waveform, the classifier should preserve the original I/Q input and adapt its internal representation according to channel quality.

Feature-wise Linear Modulation (FiLM) provides a natural mechanism for conditioning neural features on external information~\cite{perez2018film}. However, FiLM applies an unconstrained affine transformation, which can overwrite useful feature representations when the conditioning signal is imperfect or when the base representation is already informative. This is undesirable for AMR, where preserving subtle I/Q-derived structure is critical in the presence of noise. 

We therefore propose Bounded Residual Feature-wise Linear Modulation (BR-FiLM), a novel channel-quality-conditioned feature adaptation mechanism that adds bounded, gated residual corrections instead of replacing the original feature path. 
By preserving the primary feature pathway and tuning it through controlled residual adjustments, BR‑FiLM enhances robustness under adverse channel conditions while maintaining stable feature propagation. BR-FiLM is designed as a modular conditioning block that can be inserted into AMR networks with intermediate feature representations. We refer to the resulting AMR framework as BR-FiLMNet. In this paper, BR-FiLMNet is instantiated using an MCLDNN backbone, leveraging its effectiveness as a multi-stream CNN-RNN architecture for I/Q-based AMR.
In the primary experimental setting, true SNR is used as an oracle conditioning variable to isolate the effect of reliable SNR-driven feature adaptation. The normalized SNR is passed through a nonlinear conditioning network to produce feature-wise scaling and shifting parameters, which are regulated by learned scalar gates. BR-FiLM is applied across convolutional, recurrent, and Fully Connected (FC) layers, allowing channel-quality information to influence spatial feature extraction, temporal modeling, and final classification. 
The main contributions of this paper are summarized as follows:
\begin{itemize}
\item We propose BR-FiLM, a novel conditioning mechanism for AMR that uses normalized SNR to generate feature-wise scaling and shifting corrections. By employing a learned scalar gating mechanism with bounded $\tanh$ activations, BR-FiLM explicitly constrains the magnitude of these corrections relative to the input features. This design preserves the integrity of the original I/Q-driven representations, ensuring that the conditioning process augments, rather than overwrites, the underlying signal features.

\item We develop BR-FiLMNet, a multi-point channel-conditioning architecture, as our key contribution. It systematically integrates BR-FiLM across the convolutional, recurrent, and dense stages of an MCLDNN backbone. This design enables hierarchical, SNR-aware feature adaptation throughout the network, allowing channel-quality information to influence spatial, temporal, and decision-level representations, without requiring explicit waveform-level denoising.

\item We conduct extensive simulations on the RadioML 2016.10a dataset. Results show BR-FiLMNet improves overall mean accuracy from 61.79\% to 67.74\% and low-SNR mean accuracy from 37.12\% to 46.44\%, compared with the baseline MCLDNN. BR-FiLMNet also outperforms recent transformer-style baselines. 
\end{itemize}

The remainder of this paper is organized as follows. Section~II reviews related work. Section~III presents the proposed BR-FiLM block and BR-FiLMNet framework. Section~IV reports on the experimental results and analysis, and Section~V concludes the paper.

\section{Related Work}
\label{sec:related}
Traditional AMR methods are broadly categorized into likelihood-based and feature-based approaches~\cite{jdid2021machine,zhang2022deep}. Likelihood-based methods frame AMR as a multi-hypothesis testing problem but typically require prior knowledge of the signal or channel and are computationally intensive. Feature-based methods mitigate this by using handcrafted descriptors (e.g., higher-order statistics or spectral features), though their performance depends on expert design and often degrades under varying channel conditions.

Deep learning has become the dominant paradigm in AMR, as neural networks can learn discriminative representations directly from raw I/Q samples. Early studies employed CNNs to capture local spatial structure in I/Q or constellation-like representations~\cite{hermawan2020cnn,wu2019convolutional}, while RNNs were introduced to model temporal dependencies in signal sequences~\cite{hong2017automatic}. Hybrid architectures combine these strengths, using convolutional layers for feature extraction and recurrent layers for sequence modeling. Among them, MCLDNN stands out as a strong baseline, jointly processing the composite I/Q signal alongside separate I and Q streams to enable complementary spatiotemporal representation learning~\cite{xu2020spatiotemporal}.

More recently, attention-based and transformer-style AMR models have been studied to improve long-range dependency modeling and representation capacity~\cite{huo2026smtrans,shao2025iqformer1}. These architectures can capture global signal relationships that are difficult for purely convolutional or recurrent models to represent.

Low-SNR robustness remains a persistent challenge for DL-based AMR. Several works have explored denoising-assisted AMR through filtering or learned preprocessing~\cite{zhang2021denoising,liu2024dr2d}. While denoising can suppress additive noise, it may also remove fine waveform details useful for modulation discrimination, especially when class identity depends on subtle phase, amplitude, or envelope cues. This limitation motivates robustness mechanisms that adapt internal representations without directly reconstructing the received waveform.

FiLM provides a general mechanism for conditioning intermediate neural features through learned scale and shift parameters~\cite{perez2018film}. However, its use as an SNR-driven conditioning mechanism for AMR remains largely unexplored. The proposed BR-FiLM block addresses this gap by applying bounded, gated, residual SNR-conditioned corrections within a multi-stream AMR backbone.

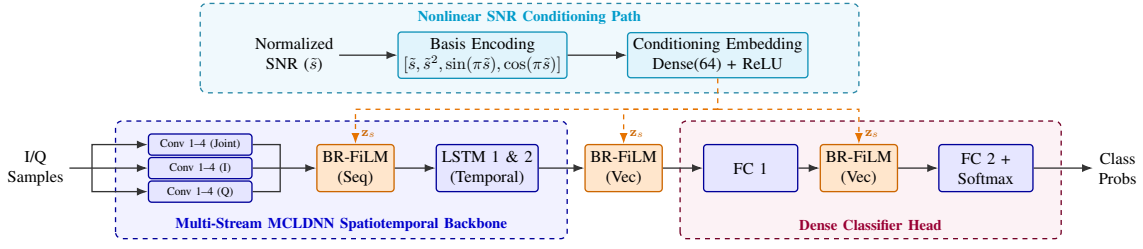
\begin{figure*}[!htbp]
\centering
\resizebox{0.84\textwidth}{!}{
\begin{tikzpicture}[
    font=\normalsize,
    >=Latex,
    mcldnn/.style={draw=blue!60!black, rounded corners=2pt, thick, align=center, minimum height=10mm, minimum width=20mm, fill=blue!10},
    mcldnn_small/.style={draw=blue!60!black, rounded corners=2pt, thick, align=center, minimum height=4.5mm, minimum width=22mm, fill=blue!10, font=\scriptsize},
    film/.style={draw=orange!80!black, rounded corners=2pt, thick, align=center, minimum height=10mm, minimum width=15mm, fill=orange!20},
    snr/.style={draw=cyan!70!black, rounded corners=2pt, thick, align=center, minimum height=9mm, minimum width=20mm, fill=cyan!10},
    arrow/.style={->, thick, color=black!80},
    line/.style={thick, color=black!80},
    condarrow/.style={->, thick, dashed, draw=orange!90!black}
]


\draw[draw=cyan!60!black, dashed, thick, rounded corners=4pt, fill=cyan!5] (1.5, 1.6) rectangle (15.5, 3.5);
\node[font=\small\bfseries, text=cyan!80!black] at (8.5, 3.2) {Nonlinear SNR Conditioning Path};

\draw[draw=blue!60!black, dashed, thick, rounded corners=4pt, fill=blue!5] (-0.3, -1.5) rectangle (9.3, 1.0);
\node[font=\small\bfseries, text=blue!80!black] at (4.5, -1.2) {Multi-Stream MCLDNN Spatiotemporal Backbone};

\draw[draw=purple!60!black, dashed, thick, rounded corners=4pt, fill=purple!5] (11.7, -1.5) rectangle (19.8, 1.0);
\node[font=\small\bfseries, text=purple!80!black] at (15.75, -1.2) {Dense Classifier Head};

\node[align=center] (I/Q) at (-2.0, 0) {I/Q\\Samples};

\node[mcldnn_small] (conv_j) at (1.5, 0.5) {Conv 1--4 (Joint)};
\node[mcldnn_small] (conv_i) at (1.5, 0) {Conv 1--4 (I)};
\node[mcldnn_small] (conv_q) at (1.5, -0.5) {Conv 1--4 (Q)};

\node[film] (f1) at (4.8, 0) {BR-FiLM\\(Seq)};
\node[mcldnn] (lstm) at (7.6, 0) {LSTM 1 \& 2\\(Temporal)};
\node[film] (f2) at (10.5, 0) {BR-FiLM\\(Vec)};

\node[mcldnn] (fc1) at (13.2, 0) {FC 1};
\node[film] (f3) at (15.5, 0) {BR-FiLM\\(Vec)};
\node[mcldnn] (out) at (18.2, 0) {FC 2 +\\Softmax};
\node[align=center] (preds) at (21.0, 0) {Class\\Probs};

\node[align=center] (snr) at (3.5, 2.4) {Normalized\\SNR ($\tilde{s}$)};
\node[snr] (basis) at (7.5, 2.4) {Basis Encoding\\$[\tilde{s}, \tilde{s}^2, \sin(\pi \tilde{s}), \cos(\pi \tilde{s})]$};
\node[snr] (dense) at (12.5, 2.4) {Conditioning Embedding\\Dense(64) + ReLU};

\coordinate (split_I/Q) at (-0.8, 0);
\draw[line] (I/Q.east) -- (split_I/Q);
\draw[arrow] (split_I/Q) |- (conv_j.west);
\draw[arrow] (split_I/Q) |- (conv_q.west);
\draw[arrow] (split_I/Q) -- (conv_i.west);

\coordinate (merge_conv) at (3.2, 0);
\draw[line] (conv_j.east) -| (merge_conv);
\draw[line] (conv_q.east) -| (merge_conv);
\draw[line] (conv_i.east) -- (merge_conv);
\draw[arrow] (merge_conv) -- (f1.west);

\draw[arrow] (f1) -- (lstm);
\draw[arrow] (lstm) -- (f2);
\draw[arrow] (f2) -- (fc1);
\draw[arrow] (fc1) -- (f3);
\draw[arrow] (f3) -- (out);
\draw[arrow] (out) -- (preds);

\draw[arrow] (snr) -- (basis);
\draw[arrow] (basis) -- (dense);

\coordinate (route_drop) at (12.5, 1.25);
\draw[line, dashed, draw=orange!90!black] (dense.south) -- (route_drop);

\draw[condarrow] (route_drop) -| (f1.north) node[pos=0.85, right, font=\footnotesize, text=orange!80!black] {$\mathbf{z}_s$};
\draw[condarrow] (route_drop) -| (f2.north) node[pos=0.85, right, font=\footnotesize, text=orange!80!black] {$\mathbf{z}_s$};
\draw[condarrow] (route_drop) -| (f3.north) node[pos=0.85, right, font=\footnotesize, text=orange!80!black] {$\mathbf{z}_s$};

\end{tikzpicture}
}
\caption{Macro-level overview of BR-FiLMNet. The I/Q stream extracts spatial and temporal features through the multi-stream MCLDNN backbone, while the SNR stream produces a shared channel-quality embedding. The proposed BR-FiLM blocks use this embedding to generate bounded residual corrections at convolutional, recurrent, and Fully Connected (FC) stages before final modulation classification.}
\label{fig:brfilm_architecture}
\vspace{-3mm}
\end{figure*}

\section{Proposed BR-FiLMNet Architecture}
\label{sec:model}
The proposed BR-FiLM block is a channel-quality-conditioned feature adaptation mechanism for AMR. Rather than modifying the received waveform through preprocessing or denoising, BR-FiLM preserves the original I/Q representation and adapts intermediate features through bounded residual conditioning. The resulting AMR framework is referred to as BR-FiLMNet.

As shown in Fig.~\ref{fig:brfilm_architecture}, BR-FiLMNet contains two coupled information streams. The I/Q stream passes through the multi-stream MCLDNN backbone to produce convolutional and recurrent feature representations. In parallel, the normalized SNR is mapped through a nonlinear conditioning path to produce a shared channel-quality embedding. This separation keeps modulation recognition driven by I/Q-derived signal features, while allowing channel-quality information to adapt these features through controlled conditioning. The SNR embedding does not classify the modulation directly; instead, it parameterizes BR-FiLM modules that inject bounded feature-wise corrections after convolutional extraction, recurrent modeling, and the first Fully Connected (FC) layer. The classifier head then operates on the conditioned representation. Although this work instantiates BR-FiLM on MCLDNN, the conditioning block is not tied to a specific backbone and can be inserted into other AMR networks with intermediate feature representations.

\subsection{Signal Representation}
The received complex baseband signal segment of length $T$ is denoted as: $\mathbf{r} = [r_1,r_2,\ldots,r_T] \in \mathbb{C}^{T}$.
For the RadioML 2016.10a setup used in this work, $T=128$. The complex signal is represented using its in-phase and quadrature components as
\begin{equation}
\mathbf{x} =
\begin{bmatrix}
\Re(\mathbf{r}) \\
\Im(\mathbf{r})
\end{bmatrix}
\in \mathbb{R}^{2\times T}.
\end{equation}
In addition to the joint I/Q tensor $\mathbf{x}$, two single-channel streams are formed:
\begin{equation}
\mathbf{x}_I = \Re(\mathbf{r}) \in \mathbb{R}^{T\times 1}, \qquad
\mathbf{x}_Q = \Im(\mathbf{r}) \in \mathbb{R}^{T\times 1}.
\end{equation}
The joint I/Q stream preserves cross-channel phase and amplitude relationships, while the separate I and Q streams allow the network to learn channel-specific temporal patterns.

Let $s_{\mathrm{dB}}$ denote the SNR value associated with the received segment. In the oracle setting considered for the main model, this SNR is provided as a channel-quality condition. It is normalized to the bounded interval $[-1,1]$ by
\begin{equation}
\tilde{s} = 2\frac{s_{\mathrm{dB}}-s_{\min}}{s_{\max}-s_{\min}}-1,
\label{eq:snr_norm}
\end{equation}
where $s_{\min}=-20$ dB and $s_{\max}=18$ dB in the dataset. The normalized variable $\tilde{s}$ is used only to condition the internal feature representations; it is not used as a standalone classifier.

\subsection{MCLDNN Multi-Stream Backbone}
The base feature extractor follows the MCLDNN design, which processes the joint I/Q tensor and the isolated I and Q streams through separate convolutional branches. Let
\begin{equation}
\begin{aligned}
\mathbf{f}_{I/Q} &= \Phi_{I/Q}(\mathbf{x}_f),\\
\mathbf{f}_{I} &= \Phi_{I}(\mathbf{x}_I), \\ \mathbf{f}_{Q} &= \Phi_{Q}(\mathbf{x}_Q),
\end{aligned}
\end{equation}
where $\Phi_{I/Q}(\cdot)$ denotes the 2D convolutional branch applied to the joint I/Q representation, and $\Phi_I(\cdot)$ and $\Phi_Q(\cdot)$ denote causal 1D convolutional branches applied to the I and Q streams, respectively.

The branch outputs are concatenated and processed by convolutional fusion layers: $
\mathbf{F}_{c} = \Phi_{F}\left([\mathbf{f}_{I/Q};\mathbf{f}_I;\mathbf{f}_Q]\right),$ where $[\cdot;\cdot]$ denotes feature concatenation and $\Phi_F(\cdot)$ denotes the post-concatenation convolutional fusion stack. This fused representation captures local spatial and cross-channel relationships among the I/Q components.

After convolutional fusion, the resulting feature map is reshaped into a temporal sequence $
\mathbf{G}=[\mathbf{g}_1,\mathbf{g}_2,\ldots,\mathbf{g}_L],
$ which is processed by two stacked Long Short-Term Memory (LSTM) layers:
\begin{equation}
\begin{aligned}
\mathbf{h}^{(1)}_t &= \mathrm{LSTM}_1(\mathbf{g}_t,\mathbf{h}^{(1)}_{t-1}), \\
\mathbf{h}^{(2)}_t &= \mathrm{LSTM}_2(\mathbf{h}^{(1)}_t,\mathbf{h}^{(2)}_{t-1}).
\end{aligned}
\end{equation}
The final spatiotemporal latent representation is taken from the last temporal step as $\mathbf{h} = \mathbf{h}^{(2)}_L$.

\vspace{-0.5mm}
\subsection{Nonlinear SNR Conditioning Path}
\vspace{-0.5mm}

A central limitation of directly feeding a raw scalar SNR into a conditioning layer is that a single scalar projection provides limited capacity to represent non-monotonic SNR-dependent behavior. In AMR, the effect of channel quality is not strictly monotonic across all feature levels: low-SNR, transition-SNR, and high-SNR regimes may require different feature corrections. To provide a richer conditioning signal, the normalized SNR is expanded into a nonlinear basis:
$
\mathbf{b}(\tilde{s}) = 
\left[
\tilde{s}, \ 
\tilde{s}^{2}, \ 
\sin(\pi\tilde{s}), \ 
\cos(\pi\tilde{s})
\right].
$
This basis is mapped into a shared 64-dimensional condition embedding through a lightweight MLP:
\begin{equation}
\begin{aligned}
\mathbf{u}_s &= \rho(\mathbf{W}_1\mathbf{b}(\tilde{s})+\mathbf{b}_1), \\
\mathbf{z}_s &= \rho(\mathbf{W}_2\mathbf{u}_s+\mathbf{b}_2),
\end{aligned}
\end{equation}
where $\rho(\cdot)$ denotes the ReLU activation. The resulting embedding $\mathbf{z}_s$ is shared by all BR-FiLM injection sites.

\vspace{-0.5mm}
\subsection{Bounded Residual FiLM Block}
\begin{figure}[!t]
\centering
\begin{tikzpicture}[
  scale=0.74,
  transform shape,
  font=\small,
  >=Latex,
  input/.style={draw, rounded corners=2pt, thick, align=center,
      minimum height=7.5mm, minimum width=15mm, fill=green!18},
  cond/.style={draw, rounded corners=2pt, thick, align=center,
      minimum height=7.5mm, minimum width=18mm, fill=green!15},
  head/.style={draw, rounded corners=2pt, thick, align=center,
      minimum height=7.5mm, minimum width=16mm, fill=orange!20},
  block/.style={draw, rounded corners=2pt, thick, align=center,
      minimum height=8mm, minimum width=17mm, fill=blue!12},
  feat/.style={draw, rounded corners=2pt, thick, align=center,
      minimum height=7.5mm, minimum width=15mm, fill=purple!15},
  op/.style={draw, circle, thick, minimum size=5.2mm, inner sep=0pt,
      fill=yellow!25},
  outblock/.style={draw, rounded corners=2pt, thick, align=center,
      minimum height=7.5mm, minimum width=15mm, fill=red!15},
  line/.style={-Latex, thick},
  dline/.style={-Latex, thick, dashed}
]

\node[input] (snr) at (0,0) {Norm. SNR\\$\tilde{s}$};
\node[cond] (basis) at (2.55,0) {Basis\\$[\tilde{s},\tilde{s}^{2},\sin,\cos]$};
\node[cond] (emb) at (5.25,0) {SNR embed.\\Dense 64};

\node[head] (gamma) at (1.25,-1.65) {$\gamma$ head\\$\tanh$};
\node[head] (beta)  at (3.45,-1.65) {$\beta$ head\\$\tanh$};
\node[head] (gate)  at (5.65,-1.65) {$g$ head\\$\sigma$};

\node[block] (scaleg) at (1.25,-3.10) {Scale\\$0.25\gamma$};
\node[block] (scaleb) at (3.45,-3.10) {Shift\\$0.25\beta$};
\node[block] (gbox)   at (5.65,-3.10) {Gate\\$g$};

\node[feat] (z) at (-0.35,-4.65) {Input\\$\mathbf{z}$};
\node[op] (mul1) at (1.25,-4.65) {$\odot$};
\node[op] (sum1) at (3.45,-4.65) {$+$};
\node[op] (mul2) at (5.65,-4.65) {$\times$};
\node[op] (sum2) at (7.05,-4.65) {$+$};
\node[outblock] (zout) at (8.55,-4.65) {Output\\$\mathbf{z}'$};

\draw[line] (snr.east) -- ++(0.25,0) -- (basis.west);
\draw[line] (basis.east) -- ++(0.25,0) -- (emb.west);

\coordinate (filmSplit) at (5.25,-0.80);
\draw[line] (emb.south) -- (filmSplit);
\draw[line] (filmSplit) -| (gamma.north);
\draw[line] (filmSplit) -| (beta.north);
\draw[line] (filmSplit) -| (gate.north);

\draw[dline] (gamma.south) -- (scaleg.north);
\draw[dline] (beta.south) -- (scaleb.north);
\draw[dline] (gate.south) -- (gbox.north);

\draw[dline] (scaleg.south) -- (mul1.north);
\draw[dline] (scaleb.south) -- (sum1.north);
\draw[dline] (gbox.south) -- (mul2.north);

\draw[line] (z.east) -- (mul1.west);
\draw[line] (mul1.east) -- (sum1.west);
\draw[line] (sum1.east) -- (mul2.west);
\draw[line] (mul2.east) -- (sum2.west);
\draw[line] (sum2.east) -- (zout.west);

\draw[line] (z.south) -- ++(0,-0.62) -| (sum2.south);

\end{tikzpicture}
\vspace{1mm}
\caption{Structure of the proposed BR-FiLM conditioning block. The shared SNR embedding generates bounded scale, shift, and gate parameters, which apply a gated residual correction to the input feature:
$\mathbf{z}'=\mathbf{z}+g(\tilde{s})\left(0.25\,\gamma(\tilde{s})\odot\mathbf{z}+0.25\,\beta(\tilde{s})\right)$.}
\label{fig:micro_architecture}
\end{figure}
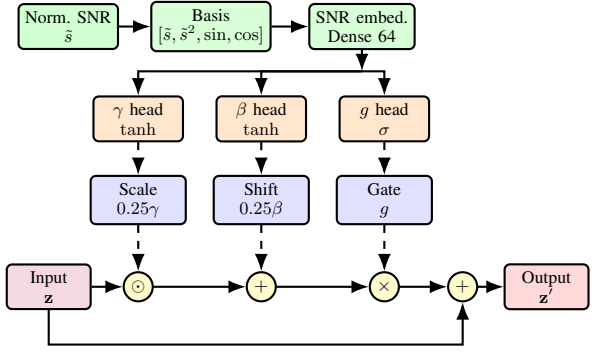

\vspace{-0.5mm}
The internal structure of the proposed BR-FiLM block is shown in Fig.~\ref{fig:micro_architecture}. At each BR-FiLM injection site \(l\), the shared SNR embedding \(\mathbf{z}_s\) generates feature-wise scale and shift vectors and a scalar residual gate:
\begin{equation}
\begin{aligned}
  \boldsymbol{\gamma}_l(\tilde{s}) &= \tanh(\mathbf{W}_{\gamma,l}\mathbf{z}_s+\mathbf{b}_{\gamma,l}), \\
\boldsymbol{\beta}_l(\tilde{s})  &= \tanh(\mathbf{W}_{\beta,l}\mathbf{z}_s+\mathbf{b}_{\beta,l}), \\
g_l(\tilde{s}) &= \sigma(\mathbf{w}_{g,l}^{T}\mathbf{z}_s+b_{g,l}).
\end{aligned}
\end{equation}
Here, \(\boldsymbol{\gamma}_l,\boldsymbol{\beta}_l\in\mathbb{R}^{d_l}\) are feature-wise, while \(g_l(\tilde{s})\in(0,1)\) is a scalar gate for site \(l\) and each sample. For sequence features, \(\boldsymbol{\gamma}_l\) and \(\boldsymbol{\beta}_l\) are broadcast across time.

Given an intermediate feature \(\mathbf{a}_l\), BR-FiLM applies
\begin{equation}
\mathbf{a}'_l
=
\mathbf{a}_l
+
g_l(\tilde{s})
\left(
\alpha\,\boldsymbol{\gamma}_l(\tilde{s})\odot\mathbf{a}_l
+
\alpha\,\boldsymbol{\beta}_l(\tilde{s})
\right),
\label{eq:brfilm}
\end{equation}
where \(\alpha=0.25\). Since \(|\gamma_{l,j}|,|\beta_{l,j}|\leq1\) and \(g_l\in[0,1]\), each feature satisfies
\begin{equation}
|a'_{l,j}-a_{l,j}|\leq0.25(|a_{l,j}|+1).
\end{equation}
Thus, BR-FiLM bounds the relative scaling and additive shift while preserving the original feature path. Unlike classic FiLM, which directly transforms features as \(\boldsymbol{\gamma}\odot\mathbf{a}+\boldsymbol{\beta}\), BR-FiLM only adds a controlled residual correction. If SNR conditioning is not useful for a given sample or layer, the learned gate can reduce the correction and fall back toward the unconditioned MCLDNN pathway.

\subsection{Multi-Point BR-FiLM Injection}
BR-FiLMNet applies BR-FiLM at three abstraction levels. First, BR-FiLM is applied after the convolutional fusion stage, before recurrent modeling:
\begin{equation}
\mathbf{F}'_{c}=\mathrm{BR\text{-}FiLM}_{c}(\mathbf{F}_{c},\tilde{s}).
\end{equation}
This allows SNR conditioning to adjust the feature sequence seen by the LSTM layers.
Second, BR-FiLM is applied to the recurrent latent representation:
\begin{equation}
\mathbf{h}'=\mathrm{BR\text{-}FiLM}_{r}(\mathbf{h},\tilde{s}).
\end{equation}
This allows the model to adapt the temporal summary according to the channel-quality regime.
Third, BR-FiLM is applied after the first fully connected layer:
\begin{equation}
\mathbf{q}'=\mathrm{BR\text{-}FiLM}_{f}(\mathbf{q},\tilde{s}),
\end{equation}
where $\mathbf{q}$ denotes the first dense feature vector. This final injection adjusts the classifier representation immediately before the output head.

The three injection sites use the same SNR embedding $\mathbf{z}_s$, but each site has its own $\boldsymbol{\gamma}$, $\boldsymbol{\beta}$, and gate projection heads. Therefore, the network can learn different SNR-dependent corrections for convolutional evidence, temporal summaries, and classifier-level features.

\subsection{Classification and Training Objective}
The final classifier maps the conditioned dense representation to a posterior distribution over $C$ modulation classes:
\begin{equation}
\hat{\mathbf{y}} = \mathrm{softmax}\left(f_c(\mathbf{q}')\right),
\end{equation}
where $\hat{\mathbf{y}}\in\mathbb{R}^{C}$. The model is trained end-to-end using a mini-batch categorical cross-entropy loss objective, defined as:
\begin{equation}
\mathcal{L}_{CE} = -\frac{1}{B} \sum_{i=1}^{B} \sum_{c=1}^{C} y_{i,c} \log \hat{y}_{i,c}
\label{eq:ce_loss}
\end{equation}
where $B$ is the mini-batch size, $C$ is the total number of modulation classes, $y_{i,c}$ is the binary ground-truth indicator for the $c$-th class of the $i$-th sample, and $\hat{y}_{i,c}$ is the corresponding predicted probability output by the softmax layer.

The complete mapping can be summarized as
\begin{equation}
\hat{\mathbf{y}} = F_{\theta}(\mathbf{x},\mathbf{x}_I,\mathbf{x}_Q,\tilde{s}),
\end{equation}
where $\theta$ contains the MCLDNN backbone, BR-FiLM conditioning path, and classifier parameters. BR-FiLMNet keeps modulation recognition driven by I/Q-derived features while using SNR only as a bounded channel-quality conditioning signal.

\section{Results and Discussion}
\label{sec:res}

\subsection{Dataset and Training Setup}
We evaluate the proposed BR-FiLMNet model on the RadioML 2016.10a\cite{o2016radio} benchmark dataset, which is widely used for AMR research. The dataset contains 220,000 signal samples from 11 modulation classes over 20 SNR levels ranging from $-20$\,dB to $18$\,dB in 2\,dB steps. Each sample consists of 128 complex I/Q samples, represented by 128 in-phase and 128 quadrature components. The modulation classes include BPSK, QPSK, 8PSK, QAM16, QAM64, CPFSK, GFSK, PAM4, WBFM, AM-SSB, and AM-DSB.

Each modulation–SNR block is split into 600 training, 200 validation, and 200 testing samples, yielding 132,000 training, 44,000 validation, and 44,000 testing instances. Overall mean accuracy is computed across the balanced SNR grid, while low-SNR accuracy is evaluated for $\mathrm{SNR} \leq 0$ dB.

BR-FiLMNet is trained using categorical cross-entropy with the Adam optimizer, incorporating phase-rotation augmentation, mild regularization, and exponential moving average (EMA) stabilization of model weights. 
All models are implemented in TensorFlow 2.20 and evaluated on an Ubuntu 22.04 system with an NVIDIA V100 GPU and dual Intel Xeon Gold 6146 CPUs.

\subsection{Simulation Results}
\begin{figure}[!t]
\centering
\includegraphics[width=\linewidth]{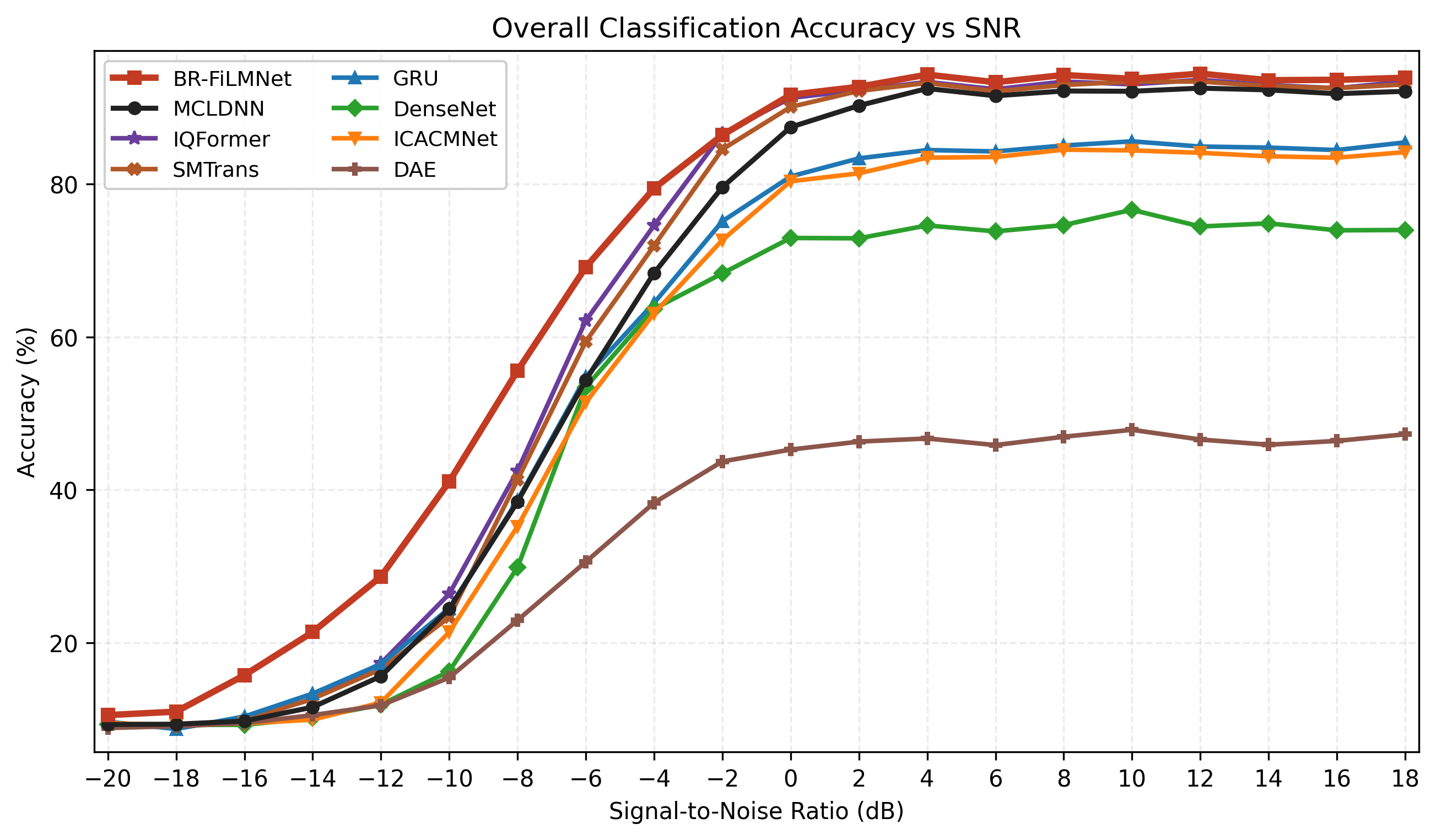}
\caption{Classification accuracy versus SNR for BR-FiLMNet and baseline models.}
\label{fig:acc_snr_main}
\end{figure}

Fig.~\ref{fig:acc_snr_main} compares BR-FiLMNet with classical neural baselines including DAE~\cite{DAE}, DenseNet~\cite{Densenet}, Gated Recurrent Unit (GRU)~\cite{hong2017automatic}, IC-ACMNet~\cite{hermawan2020cnn}, and MCLDNN~\cite{xu2020spatiotemporal}, as well as recent transformer-style baselines SMTrans and I/Q Former~\cite{huo2026smtrans,shao2025iqformer1}, across the full SNR range under the same local split protocol. BR-FiLMNet performs best in the low-SNR and transition-SNR regions while maintaining competitive high-SNR accuracy. Compared with MCLDNN, BR-FiLMNet improves overall mean accuracy by 5.95 percentage points and low-SNR mean accuracy by 9.32 percentage points. Compared with I/QFormer, BR-FiLMNet improves overall mean accuracy by 3.79 percentage points and low-SNR mean accuracy by 6.26 percentage points, indicating that bounded residual SNR conditioning is most useful when modulation structure is partially obscured by noise.

Since both models of BR-FiLMNet and MCLDNN are evaluated on the same test examples, we also conduct paired significance testing over the 44,000-example test set to verify that the gains are not due to test-set variation. Statistical results show that BR-FiLMNet improves overall accuracy over MCLDNN by 5.95\%, with a 95\% confidence interval of $[5.62, 6.28]$ and McNemar test $p = 3.134 \times 10^{-251}$. In the low-SNR region, the gain is 9.32\%, with a 95\% confidence interval of $[8.72, 9.90]$ and McNemar test $p = 4.622 \times 10^{-210}$. These results indicate that the observed improvement is statistically robust, rather than merely a fluctuation arising from the particular test split.

\begin{figure}[!b]
\centering
\includegraphics[width=\linewidth]{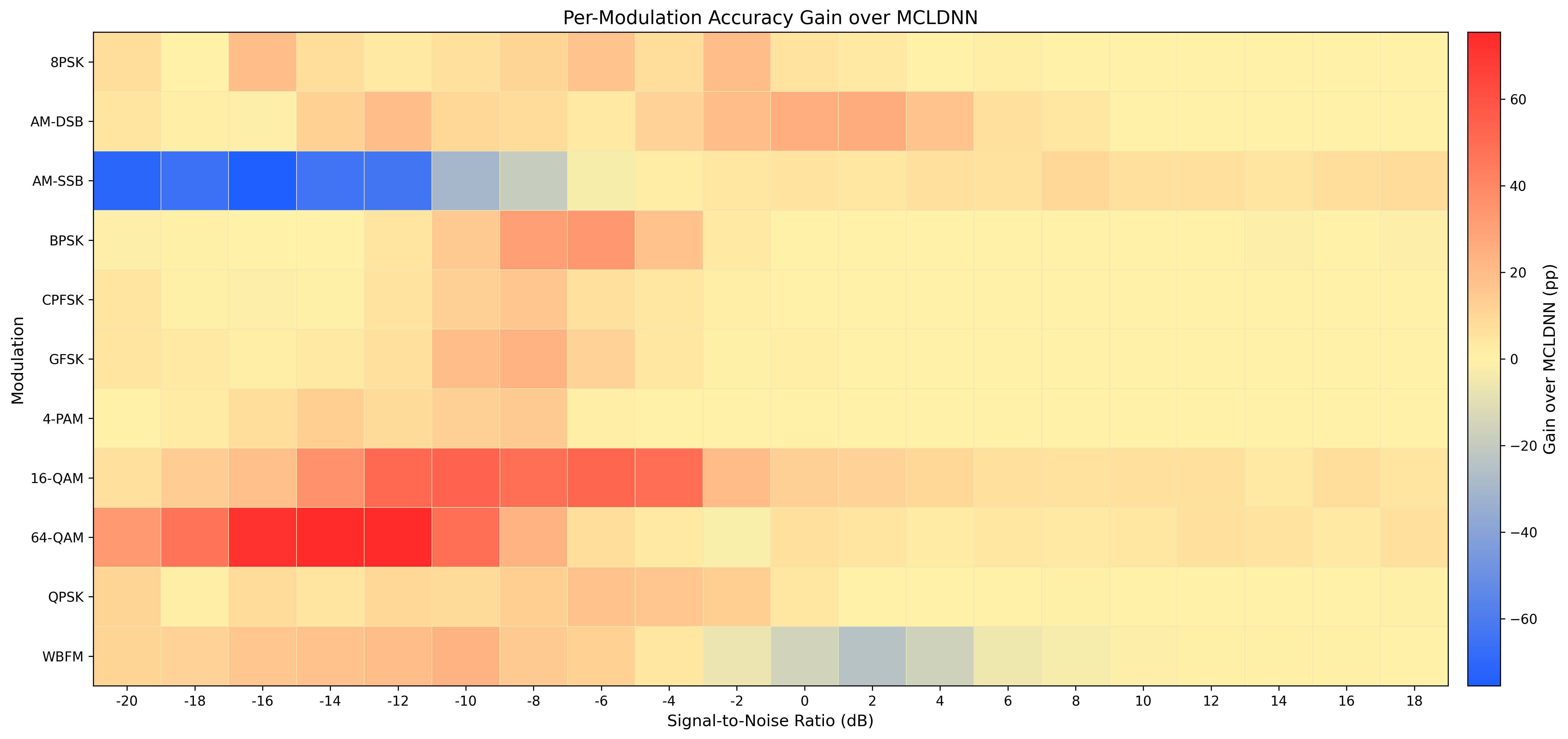}
\caption{Per-modulation accuracy gain of BR-FiLMNet over MCLDNN across SNR levels.}
\label{fig:gain_heatmap}
\end{figure}
Fig.~\ref{fig:gain_heatmap} presents the per-modulation accuracy gain of BR-FiLMNet over MCLDNN across SNR levels. The largest gains appear in the low-to-mid SNR transition region, where the classifier begins to recover modulation structure from heavy noise. The proposed BR-FiLMNet yields substantial improvements across several digital modulation classes, indicating that bounded residual conditioning helps preserve and emphasize modulation-discriminative phase and amplitude patterns when the raw I/Q representation is corrupted by noise. At the same time, the gains are not uniform across all classes. WBFM remains a difficult class, indicating that continuous analog waveform structure remains challenging for short-window I/Q-based recognition.

\begin{figure}[!t]
\centering
\includegraphics[width=\linewidth]{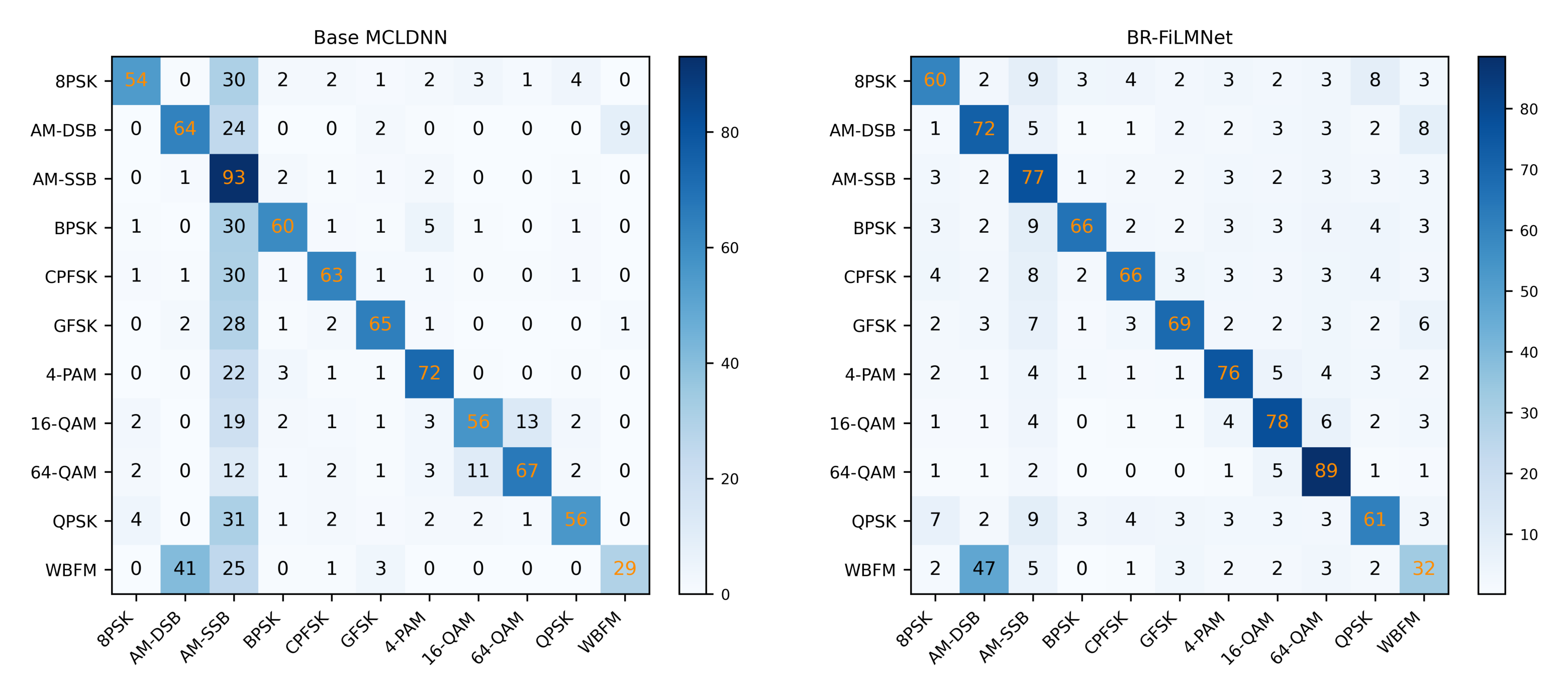}
\caption{Row-normalized confusion matrices for MCLDNN and BR-FiLMNet. Rows correspond to true classes and columns correspond to predicted classes.}
\label{fig:confusion}
\end{figure}

Fig.~\ref{fig:confusion} reports the row-normalized confusion matrices for MCLDNN and BR-FiLMNet. Compared with the MCLDNN, BR-FiLMNet produces a cleaner diagonal structure for many modulation classes, indicating improved class separability. This diagnostic view is important because the mean accuracy alone does not reveal which modulation families benefit most from conditioning or where the remaining errors occur. The remaining confusion among selected analog and frequency-modulated classes reflects the fact that BR-FiLM improves channel-quality adaptation but does not fully eliminate the intrinsic difficulty of separating all modulation families from short I/Q bursts.

\begin{figure}[!htbp]
\centering
\includegraphics[width=\linewidth]{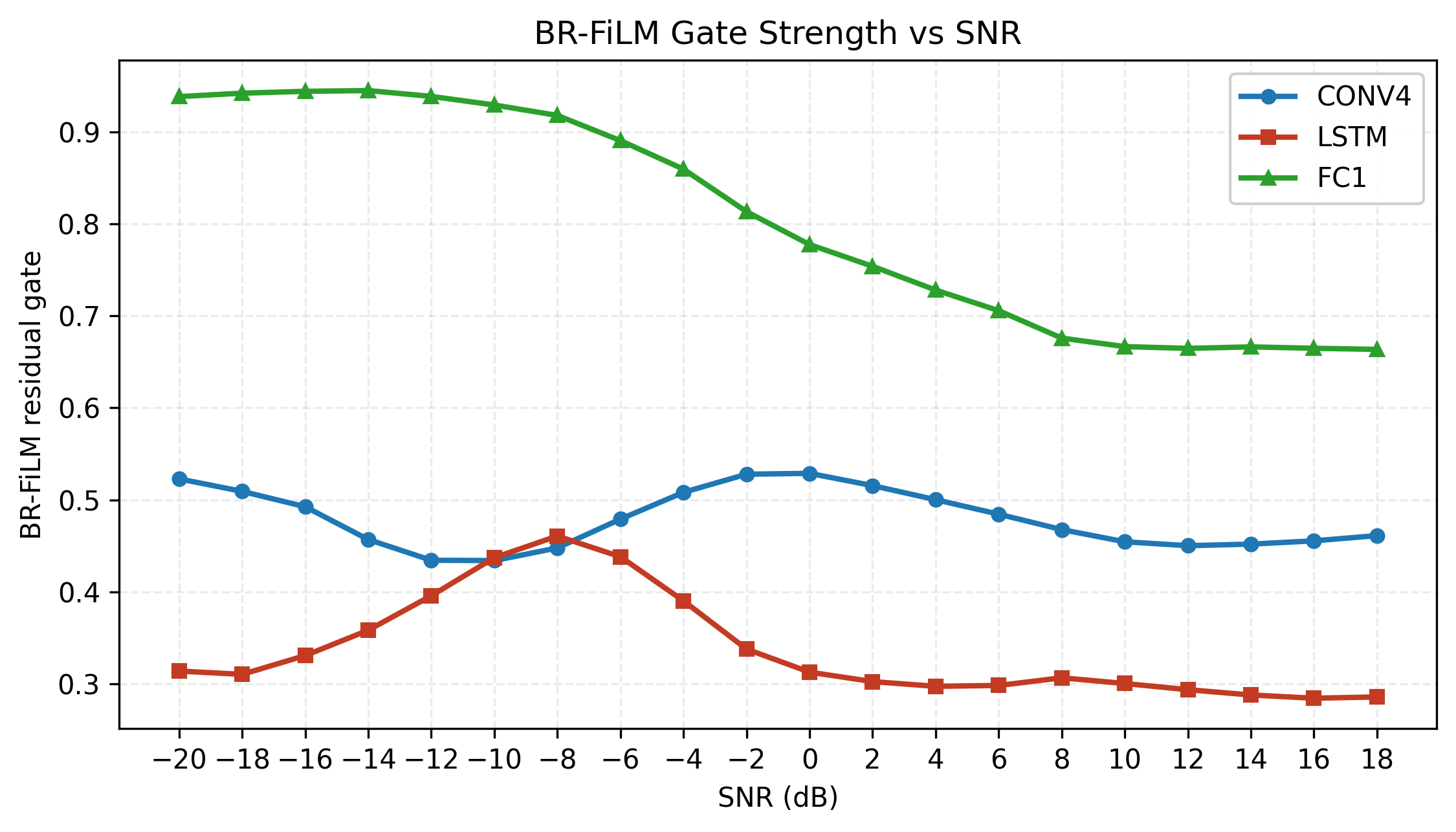}
\caption{BR-FiLM scalar gate strength across SNR levels. Each gate is sample-wise and site-specific, and is broadcast across feature dimensions.}
\label{fig:brfilm_gate}
\end{figure}

Beyond accuracy, we analyze the learned BR-FiLM gates in Fig.~\ref{fig:brfilm_gate}. The three injection sites exhibit different SNR-dependent behavior, indicating that BR-FiLMNet does not apply channel-quality conditioning uniformly across the network. The FC1 gate remains the strongest across most SNR values, suggesting that classifier-level adaptation plays a dominant role, while the convolutional and recurrent gates vary more across the low- and transition-SNR regimes. This supports the interpretation that BR-FiLM is not simply adding parameters but rather learning depth-dependent conditioning behavior.

We also evaluate computational efficiency and deployment feasibility of BR-FiLMNet compared with the MCLDNN baseline. Table~\ref{tab:complexity} reports the parameter count and batch-1 GPU inference latency of both models. BR-FiLMNet adds only a modest overhead to MCLDNN, increasing the parameter count by approximately 51k and latency by 0.65 ms per 128-sample burst.
\begin{table}[!h]
\centering
\caption{Complexity comparison of MCLDNN and BR-FiLMNet.}
\label{tab:complexity}
\begin{tabular}{lcc}
\hline
\textbf{Model} & \textbf{Params.} & \textbf{GPU Latency (ms)} \\
\hline
MCLDNN & 405,175 & 4.55 \\
BR-FiLMNet & 456,130 & 5.20 \\
\hline
\end{tabular}
\vspace{-1mm}
\end{table}

\begin{figure}[!htbp]
\centering
\includegraphics[width=\linewidth]{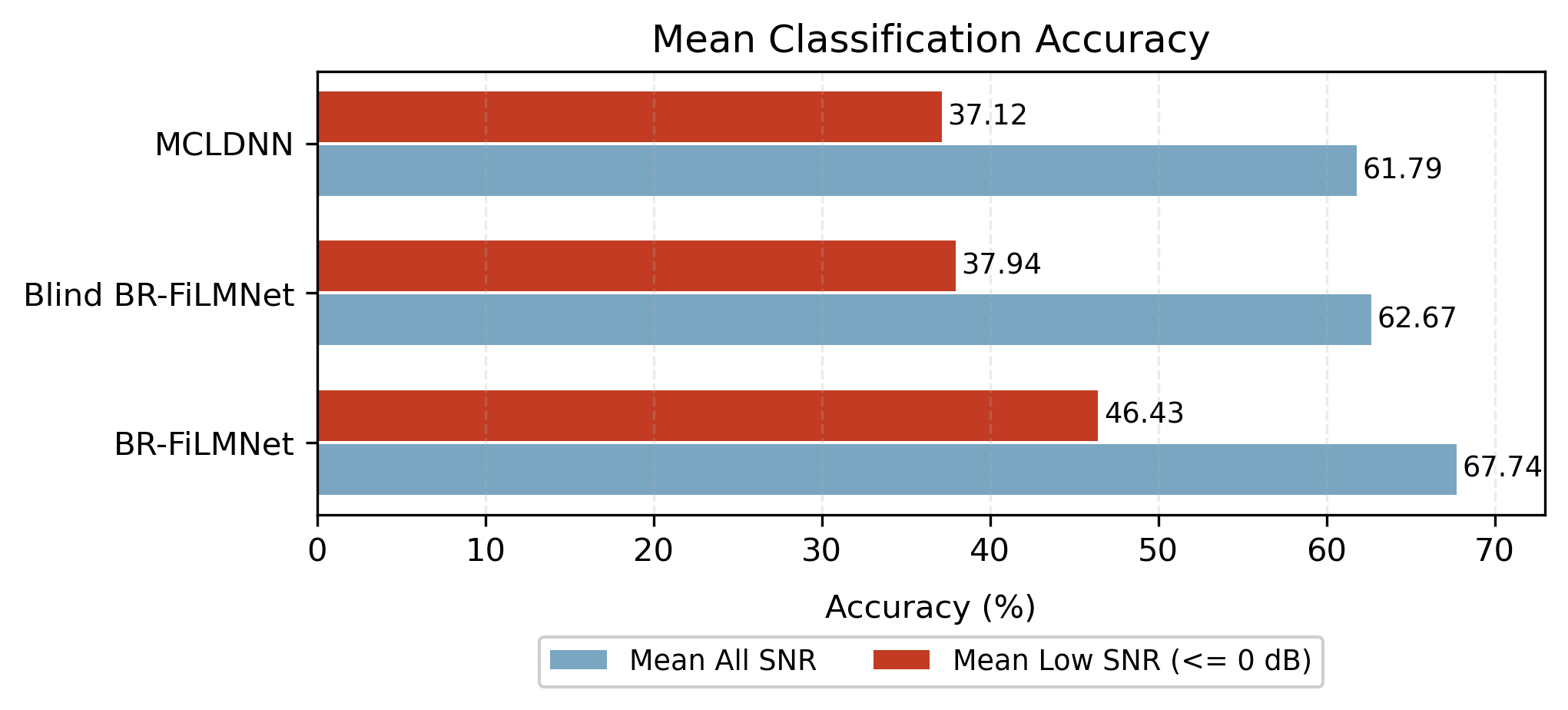}
\caption{Mean classification accuracy across MCLDNN, blind BR-FiLMNet and FiLMNet for all SNR values and for the low-SNR region $(\mathrm{SNR}\leq0~\mathrm{dB})$.}
\label{fig:mean_acc_summary}
\end{figure}

To further evaluate the practical deployment of BR-FiLMNet, we separate the effectiveness of the BR-FiLM mechanism from the distinct challenge of estimating channel quality. We first test an oracle BR-FiLMNet model using true SNR, which removes SNR-estimation error and directly measures the benefit of reliable SNR-aware feature modulation. Since true SNR may not be available in deployment, we also evaluate a blind BR-FiLMNet variant. In this variant, the true SNR input is replaced by an auxiliary I/Q-derived quality estimator that predicts a scalar SNR estimate, an SNR-bin posterior, and a learned quality embedding. The estimator is pre-trained using training-set SNR labels and then fine-tuned jointly with the classifier and BR-FiLM conditioning layers; no ground-truth SNR is used during inference. 
Fig.~\ref{fig:mean_acc_summary} summarizes the overall and low-SNR mean accuracies of the oracle BR-FiLMNet model, the blind BR-FiLMNet variant and the unconditioned MCLDNN baseline. As shown in  Fig.~\ref{fig:mean_acc_summary}, the blind variant achieves 62.68\% overall accuracy and 37.94\% low-SNR accuracy. The blind variant outperforms MCLDNN yet remains below the oracle variant, suggesting that the proposed BR-FiLMNet conditioning interface is beneficial, while accurate channel-quality estimation continues to be the main bottleneck for fully blind deployment. 

\section{Conclusion}
\label{sec:con}
In this paper, we first introduced a bounded residual, channel-quality–conditioned block termed BR‑FiLM for AMR. Building on this, we developed BR‑FiLMNet, an AMR framework that integrates the BR‑FiLM block within an MCLDNN backbone. BR‑FiLM maps normalized channel-quality information through a nonlinear transformation to generate gated residual corrections across the convolutional, recurrent, and dense layers of MCLDNN, enabling noise-aware adaptation while preserving the original I/Q feature stream.
Experiments on the RadioML 2016.10a benchmark demonstrate that BR-FiLMNet outperforms the evaluated conventional and transformer-style baselines. Compared with MCLDNN, the proposed BR-FiLMNet improves overall mean accuracy from 61.79\% to 67.74\% and low-SNR mean accuracy from 37.12\% to 46.44\%. 
Future work will focus on uncertainty-aware conditioning, improved channel-quality estimation, and addressing persistent errors in challenging modulation classes such as WBFM.

\bibliographystyle{IEEEtran}

\bibliography{reference}

\end{document}